\documentclass[useAMS,usenatbib]{mn2e}

\usepackage{graphicx}
\usepackage{url}
\usepackage{subfigure}
\usepackage{latexsym}
\usepackage{epstopdf}

\title[$T_{\rm M}(z)$]{Matter temperature during cosmological recombination}

\author[Douglas Scott and Adam Moss]{Douglas Scott\thanks{E-mail:
dscott@phas.ubc.ca},
and Adam Moss\thanks{E-mail: adammoss@phas.ubc.ca} \\
\vspace*{6pt} \\
Department of Physics and Astronomy, University of British
       Columbia, Vancouver, BC, Canada V6T 1Z1 \\
}
\begin{document}
\maketitle

\begin{abstract}
The temperature of the atomic matter in the Universe is held to that of the
Cosmic Background radiation until decoupling at $z\,{\sim}\,100$.  After this
it cools faster than the radiation ($\propto(1+z)^2$ rather than $(1+z)$) and
would have fallen to about $20\,$mK today if astrophysical feedback processes
had not heated up the interglactic medium.  We show how the derivative of
the Compton coupling equation helps numerically to follow the decoupling
process.
\end{abstract}



At early times atoms are coupled to Cosmic Microwave Background (CMB) photons
through Compton scattering.  In an expanding Universe matter would `like' to
cool as $T_{\rm M}\propto (1+z)^2$, i.e.~{\it faster\/} than the radiation,
which varies with redshift $z$ as $T_{\rm R}\propto (1+z)$.  However, Compton
coupling prevents the matter cooling this rapidly until the 
partial ionization of the atoms has fallen enough that the Compton heating
timescale becomes long compared with the Hubble time.  So although cosmological
recombination is often referred to as `decoupling', the huge photon bath
prevents matter decoupling from the radiation until much later.  The CMB
`last scattering' epoch is at $z\,{\simeq}\,1100$, while matter does not
really start to cool adiabatically until $z\,{\simeq}\,300$.

Using the {\sl WMAP\/} Markov Chains to account for the variation in the
cosmological model, parameters we find that $T_{\rm M}$ would be
$(0.0215\pm0.0002)\,$Kelvin, if there had been no additional
sources of heat.  This means that the asymptotic behaviour is the same as if
the matter had instantaneously departed from the radiation at
$1+z=2.725/0.0215\simeq127$.
Of course the growth of non-linear structure at $z\,{\la}\,20$ and
subsequent feedback of gravitational and nuclear energy leads to intergalactic
medium temperatures in today's Universe which are much {\it higher\/} than the
CMB temperature.

Although we found the uncertainty on today's $T_{\rm M}$ by considering the
variation among currently acceptable cosmological model parameters, probably a
bigger uncertainty lies in the actual physics of recombination at low redshift.
Some simple algebra shows that $T_{\rm M} \propto x_{\rm e, \, f}^{2/5}$, where
$x_{\rm e, \, f}$ is the free electron fraction (normalized to hydrogen by
$x_{\rm e}\equiv n_{\rm e}/n_{\rm H}$) which `freezes out' at low redshift.
The additional uncertainty in $T_{\rm M}$ due to $x_{\rm e, \, f}$ is then
expected to be of the order 2--4\% (see for example~\citealt{CRS}). 

The explicit equation governing the kinetic
temperature of the matter (here meaning electrons plus ions plus atoms, with
dark matter being uncoupled) is given by equation (66) in \citet{SSS}.
Ignoring the negligible atomic cooling processes (Bremsstrahlung, collisions,
etc.) we have:
\begin{equation}
(1+z) {T_{\rm M}^\prime} = {(T_{\rm M}-T_{\rm R})\over H\, t_{\rm C}}
 +2T_{\rm M},
 \label{eq:TMP}
\label{eq:dTdz}
\end{equation}
where $H(z)$ is the Hubble parameter evaluated at epoch $z$ and
\begin{equation}
t_{\rm C} \equiv 
 {3 m_{\rm e} c \over 8\sigma_{\rm T} a_{\rm R}T_{\rm R}^4}
  {1+f_{\rm He}+x_{\rm e} \over x_{\rm e}}.
\label{eq:TC}
\end{equation}
Here the prime denotes differentiation with respect to redshift,
$\sigma_{\rm T}$ is the Thomson cross-section, $a_{\rm R}$ is the
radiation constant ($=8\pi^5 k^4/15c^3h^3$)  and $f_{\rm He}$ is the
fractional abundance of helium by number (assumed ionized here for simplicity, but correctly dealt with in the full recombination codes).

Clearly $T_{\rm M}$ is slightly below $T_{\rm R}$ at early epochs, with
the difference kept at just the right value for Compton heating to make the
matter track the radiation.  The small imbalance is also important because
it produces a `Compton drag' force on the matter particles.
An estimate of this temperature difference appears
to have been first been mentioned by \citet{Gamow49}.  He states (equation
20) without proof that 
\begin{equation}
{T_{\rm R}-T_{\rm M}\over T_{\rm M}} \simeq {t{\rm (years)}\over 10^{12}}.
\end{equation}

Further discussion of this temperature difference is given by
\citet{Weymann66}, whose result is
\begin{equation}
 {T_{\rm R}-T_{\rm M}\over T_{\rm R}}
 \simeq 60 {1\over x_{\rm e}} (1+z)^{-5/2},
\end{equation}
where we have converted to our notation.  This has the approximately
right redshift dependence and ionization dependence, although differences
in assumptions about the cosmological model make it difficult to compare
the coefficient.  Still, this result is essentially correct.

The difference between matter and radiation temperatures was also included
in the textbook of
\cite{Peebles71}.  He writes a version of equation~(\ref{eq:dTdz}) and states
`Because the coefficient in the last term is so
very large we get a good approximation to the solution by setting
$T_{\rm M}^\prime=0$' (converting to our notation), hence finding that
\begin{equation}
{T_{\rm R}-T_{\rm M}\over T_{\rm R}}
\simeq {3\over 2} H
 {m_{\rm e} c\over \sigma_{\rm T} a_{\rm R} T_{\rm R}^4}.
\end{equation}
This is a good order of magnitude estimate, agreeing (in essence) with
\cite{Weymann66}, but it is worth pointing out that
$T_{\rm M}^\prime = T_{\rm R}^\prime$ would be a much better
approximation than $T_{\rm M}^\prime=0$.

Let us write $T_{\rm M}=T_{\rm R}-\epsilon$ at early times, with $\epsilon$
having the dimensions of temperature and fixing
$T_{\rm R}\propto (1+z)$ at all times.  Then the solution to
equation~(\ref{eq:dTdz}) is simply
\begin{equation}
{\epsilon\over T_{\rm R}}
 = H t_{\rm C}.
\end{equation}
In the limit $x_{\rm e}\to 1$ (and ignoring helium) this is half of the
expression in Peebles (1971).  We note that the same result is obtained in a
rather different way in the Appendix of \citet{Hirata08}.

With this approximation in hand we can write down an expression for the
evolution of the matter temperature by differentiating equation~(\ref{eq:TMP}):
\begin{equation}
T_{\rm M}^\prime = {T_{\rm R}\over (1+z)} + \epsilon \left\{
 {1+f_{\rm He} \over 1+f_{\rm He}+x_{\rm e}} {x_{\rm e}^\prime\over x_{\rm e}}
  + \left[{3\over (1+z)} - {H^\prime \over H}\right] \right\}.
\label{eq:depsdz}
\end{equation}
This expression is useful for improving the numerical accuracy of the
solution to the matter temperature.
The last two terms (in square brackets) are of similar magnitude and track
each other at high redshift, hence can be combined.  The second term
depends on the derivative of the ionization fraction, and so contributes
differently as a function of redshift.  Together the derivative in
equation~(\ref{eq:depsdz}) can be used to evolve the matter temperature
to quite high accuracy until the departure of $T_{\rm M}$ from $T_{\rm R}$
stops being small.

\begin{figure}
\begin{center}
\includegraphics[width=8.5truecm]{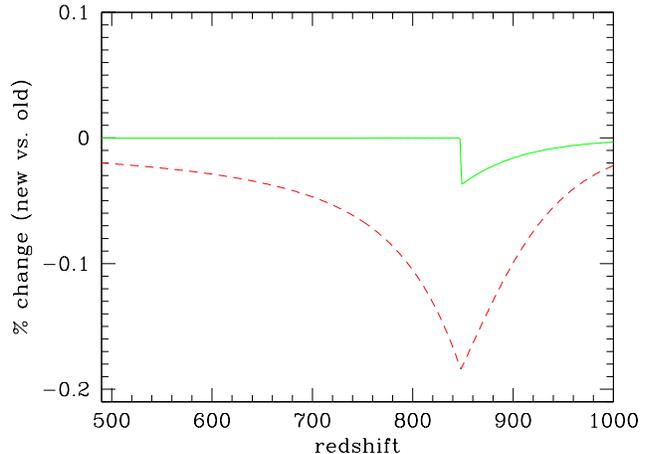}
\caption{Percentage change in the matter temperature (solid line) and the
ionization fraction (dashed line) when using the approximate derivative
before switching to the full derivative at $z\simeq850$.}
\label{fig:thefigure}
\end{center}
\end{figure}

In solving the coupled recombination equations one does not really need to
follow $T_{\rm M}$ explicitly at early times.  In the commonly used code
{\sc recfast} \citep{SSSletter,SSS,WMS} $T_{\rm M}$ is set to $T_{\rm R}$
until $H t_{\rm C}$ reaches some predefined value, and the full
equation~(\ref{eq:dTdz}) is switched on afterwards (typically at
$z\,{\simeq}\,850$).  This leads to a `glitch'
in the solution (pointed out in \citealt{Fendt08}).
We found that this glitch is easily removed by following
equation~(\ref{eq:depsdz}) instead of just $T_{\rm M}^\prime=T_{\rm M}/(1+z)$
before the switch, and then solving the full equation~(\ref{eq:dTdz})
afterwards.  The results are shown in Fig.~1.  The change in ionization
fraction, roughly $0.2 \%$ at $z=850$, leads to a $ \simeq 0.2 \%$ correction
to the CMB power spectrum $C_{\ell}$s when using  {\sc recfast}  in a Boltzmann
code. It may seem that unnecessary calculations are being carried out by
explicitly integrating the matter temperature at early times, but in fact the
integrator is already so fast that there is negligible effect on the speed
at which {\sc recfast} runs.

\section*{Acknowledgements}
This work was supported by the Natural Sciences and Engineering Research
Council of Canada and by the Canadian Space Agency.  We acknowledge the use of
the Legacy Archive for Microwave Background Data Analysis (LAMBDA).  Support
for LAMBDA is provided by the NASA Office of Space Science.

\end{document}